\documentclass{cernrep}

\usepackage{varwidth}
\usepackage{xcolor}

\begin{document}

\title{Beam Dynamics and Beam Losses -- Circular Machines}

\author{V. Kain}

\institute{CERN, Geneva, Switzerland}

\maketitle 

\begin{abstract}
A basic introduction to transverse and longitudinal beam dynamics as well as the most relevant beam loss mechanisms in circular machines will be presented in this lecture. This lecture is intended for physicists and engineers with little or no knowledge of this subject.
\end{abstract}

\section{Introduction}
A vast variety of mechanisms can lead to beam losses in accelerators. Examples are collisions in colliders, beam gas interactions, intra-beam scattering, the Touscheck effect, RF noise, collective effects, transition crossing, equipment failures and many more. Losses have an impact on perform\-ance, such as the luminosity in a collider or the brightness of the beam. Losses lead to radio-activation, which can have an impact on machine availability and maintainability. Hands-on maintainability requires radi\-ation of less than $1\UmSv/\UhZ$. High losses can cause downtime of the accelerator, due to quenches in superconducting machines, or even damage to components.

Particles are lost in the vacuum chamber if their transverse trajectory amplitudes are larger than the dimension of the vacuum chamber. It is important to understand the mechanisms that can create large amplitudes, to give input for the design of machine protection reaction times, collimators, absorbers, instrumentation, etc. The important characteristic in this respect is the number of particles lost per unit time $\Delta N/ \Delta t$.
The so-called beam lifetime $\tau$ is defined with
\begin{equation}
N(t) = N_0 \cdot \mathrm{e}^{-\frac{t}{\tau}}\ ,
\end{equation}
where $N_0$ is the initial intensity. At the accelerator design stage, such questions as, ``What is the minimum possible beam lifetime?'' and, ``What is the tolerable beam loss rate for accelerator components?'' have to be answered. This lecture will discuss some of the typical beam loss mechanisms in circular machines. To make the discussion accessible to non-accelerator physicists, a good part of the lecture will be spent on introducing the most basic concepts of accelerator physics.
\section{Principles of transverse and longitudinal beam dynamics -- synchrotrons}
This part of the lecture will only introduce the concepts required later on for the discussion of typical beam loss mechanisms. It is based on the lectures by B. Holzer, F. Tecker and O. Bruning at the CERN Accelerator School \cite{bib:CAS}. A complete introduction to the subject of accelerator physics can be found in Ref. \cite{bib:lee}.

\begin{figure}
\centering\includegraphics[width=.7\linewidth]{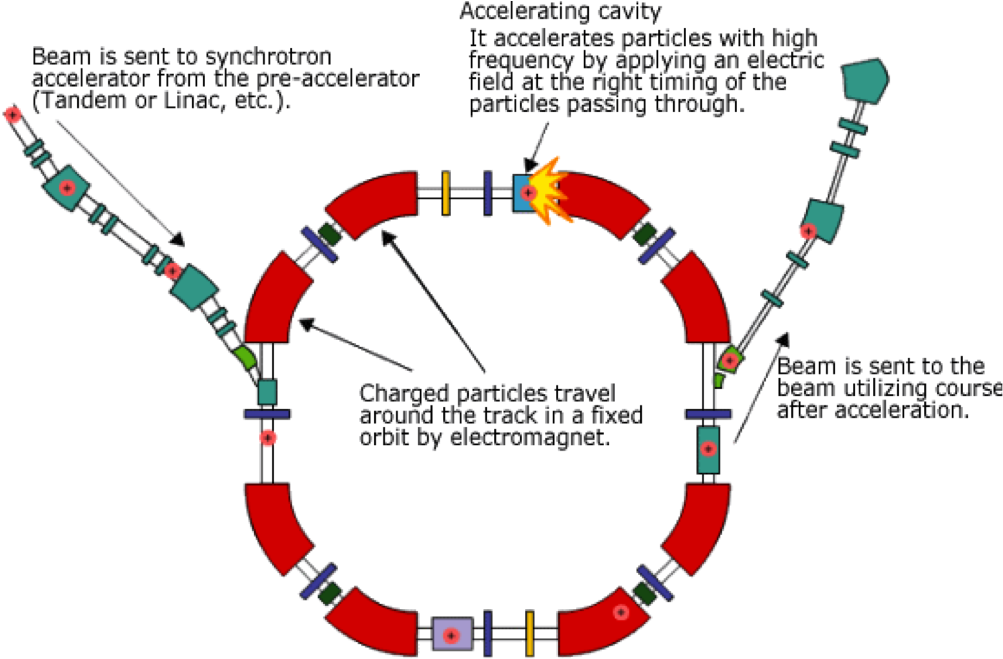}
\caption{Typical layout of synchrotron}
\label{fig:synch}
\end{figure}

A typical layout of a synchrotron is shown in Fig.\ \ref{fig:synch}. Bending magnets are used to keep the particles on the synchrotron orbit. Strong focusing from an alternating gradient lattice ensures trajectory stability over many turns. Radio-frequency accelerating structures increase the particle momentum turn by turn. Other insertions are arranged to inject or extract the beam with dedicated equipment.
\subsection{The transverse plane}
\begin{figure}
\centering\includegraphics[width=.5\linewidth]{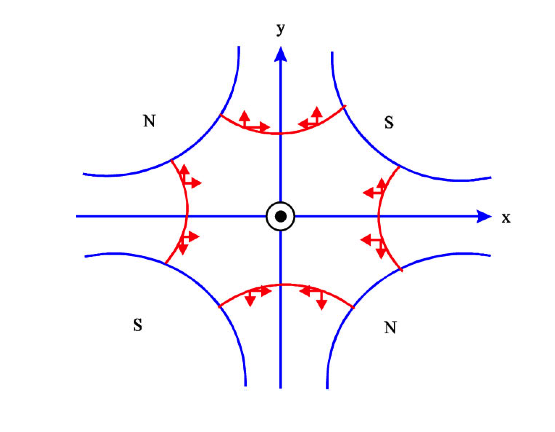}
\caption{A schematic cross-section of a quadrupole magnet. The red arrows indicate the direction of the force on the particle. A quadrupole that is focusing in the horizontal plane is defocusing in the vertical plane.}
\label{fig:quad}
\end{figure}

The trajectories of charged particles can be manipulated with electromagnetic fields via the Lorentz force:
\begin{equation}
\vec{F} = q\cdot ( \vec{E}+\vec{v}\times\vec{B})\ .
\label{eq:lorentz}\end{equation}
For relativistic particles, the effect of magnetic fields is much enhanced via the product with the velocity, and dipole fields are mainly used as guide fields. For the particles to stay on a circular orbit in a circular machine, the Lorentz force $F_\mathrm{L}$ from the magnetic field has to compensate the centrifugal force $F_{\mathrm{centr}}$.
\begin{equation}
\begin{array}{ll}
F_\mathrm{L} & = qvB\ , \\
F_{\mathrm{centr}} & = \frac{mv^2}{\rho}\ .
\end{array}
\end{equation}
From
\begin{equation}
\frac{mv^2}{\rho}=qvB\ ,
\end{equation}
the well-known relation for the product $B\rho$, the beam rigidity, follows:
\begin{equation}
\frac{p}{q}=B\rho\ .
\end{equation}
A useful formula for `back-of-the-envelope' estimates is
\begin{equation}
\frac{1}{\rho\ [\UmZ]}\approx 0.3 \frac{B\ [\UTZ]}{p\ [\UGeVcZ]}\ .
\end{equation}
Vertical dipole magnets define the design trajectory in the horizontal plane. In a beam of many par\-ticles, the trajectories of the particles will deviate from the design trajectory. Without a restoring force, the trajectories will deviate more and more until the particles are eventually lost. Quadrupole magnets provide the required restoring force. They produce a dipole field in the horizontal and vertical plane that increases as a function of the distance from the design trajectory. A schematic cross-section of a quadrupole with its field lines is given in Fig. \ref{fig:quad}. For example, the vertical field in a quadrupole will be a function of the horizontal position:
\begin{equation}
F(x) = q\cdot v\cdot B(x)\ .
\end{equation}
It depends linearly on the deviation from the design trajectory:
\begin{equation}
B_y = g\cdot x\ .
\end{equation}
The horizontal field is
\begin{equation}
B_x = g\cdot y\ .
\end{equation}
A focusing quadrupole in the horizontal plane will be defocusing in the vertical one and vice versa. The characteristic parameter of a quadrupole magnet is its gradient,
\begin{equation}
g=\frac{2\mu_0 n I}{r^2}\left[\frac{\UTZ}{\UmZ}\right]\ ,
\end{equation}
where $r$ is the distance between the quadruple centre and the pole surface. The normalized gradient is often used to define the strength of the quadrupole:
\begin{equation}
k = \frac{g}{p/e}[\UmZ^{-2}]\ .
\end{equation}

\subsubsection{Equation of motion}
\begin{figure}
\centering\includegraphics[width=.5\linewidth]{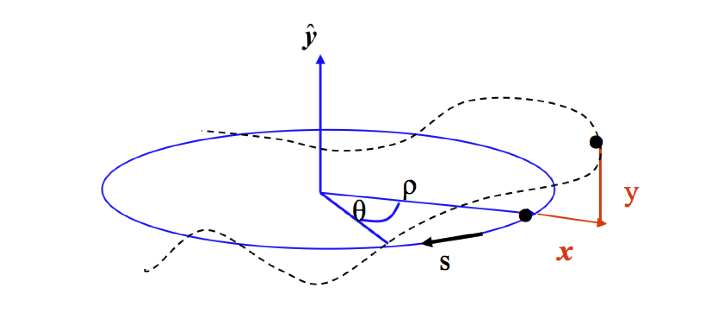}
\caption{The trajectory coordinates are given with respect to the Frenet--Serret frame, which rotates with the ideal particle around the accelerator. The ideal particle has design momentum $p_0 = m_0\gamma v$. It has coordinates $x = 0$, $y = 0$ for a certain longitudinal location $s$ lying on the design orbit.}
\label{fig:frenet_serret}
\end{figure}

To describe the particle trajectories in the synchrotron, the equation of motion has to be solved:
\begin{equation}
F_r=m\ a_r = eB_y v\ .
\end{equation}
To simplify the task, we use the Frenet--Serret coordinate system, see Fig.\  \ref{fig:frenet_serret}, and the magnetic field is expanded in a Taylor series,
\begin{equation}
B_y(x)=B_{y0} +\frac{\partial B_y}{\partial x}x
+\frac{1}{2}\frac{\partial^2 B_y}{\partial x^2}x^2
+\frac{1}{3!}\frac{\partial^3 B_y}{\partial x^3}x^3+\dots \ ,
\end{equation}
and normalized with $p/e$,
\begin{equation}
\frac{B_y(x)}{p/e}=\frac{1}{\rho}+k\ x
+\frac{1}{2} m\ x^2
+\frac{1}{3!} n\ x^3+\dots
\end{equation}
Only the terms linear in $x$ are kept:
\begin{equation}
\frac{B_y(x)}{p/e}\approx\frac{1}{\rho}+k\ x\ .
\label{eq:taylor_lin}
\end{equation}
Using the magnetic field as defined in Eq.\ (\ref{eq:taylor_lin}), the equation of motion in the horizontal plane in the Frenet--Serret frame turns out to be
\begin{equation}
x''+ x\left(\frac{1}{\rho^2}-k\right) = x''+ x K = 0\ ,
\label{eq:horizontal_diff_eq}
\end{equation}
where $x' = \mathrm{d}x/\mathrm{d}s$ and $K$ combines the focusing properties of dipoles and quadrupoles. Assuming that there are no vertical bending magnets, the equation of motion in the vertical plane becomes
\begin{equation}
 y''+ ky = 0\ .
\end{equation}
\label{eq:vertical_diff_eq}%
Around the accelerator, $K$ will not be constant, but will depend on $s$.  However, $K(s)$ will  be periodic with $L$, $K(s+L)= K(s)$, where $L$ is the lattice period. For instance, $L$ can be the circumference of the accelerator. We then have
\begin{equation}
x''(s) + K(s)x(s) =0\ .
\end{equation}
This type of equation of motion with these characteristics of non-constant but periodic restoring force is called the Hill equation, after George Hill, an astronomer of the 19th century. The general solution of the Hill equation is a quasi-harmonic oscillation:
\begin{equation}
x(s)=\sqrt{\epsilon}\sqrt{\beta (s)}\cos (\psi (s)+\phi)\ .
\end{equation}
In the case of accelerators, this quasi-harmonic oscillation is called \textit{betatron oscillation}. The amplitude and phase of the oscillation depend on the position in the ring. $\epsilon$ and $\phi$ are integration constants and depend on the initial conditions. The so-called \textit{beta function}, $\beta (s)$, is a periodic function, $\beta (s+L)~=~\beta(s)$, and is determined by the focusing properties of the lattice, i.e. quadrupole strengths. The \textit{phase advance} of the oscillation between point $ s = 0$ and point $ s$ in the lattice is
\begin{equation}
\psi (s) = \int^s_0 \frac{\mathrm{d}s}{\beta (s)}\ .
\end{equation}
Two other functions are commonly used: $\alpha (s)$ and $\gamma (s)$. They are defined as
\begin{equation}
\alpha (s) = -\frac{1}{2} \beta' (s)\ ,
\end{equation}
\begin{equation}
\gamma (s) = \frac{1 + \alpha (s)^2}{\beta (s)}\ .
\end{equation}
\subsubsection{The transport matrix}
The integration constants $\phi$ and $\epsilon$ can be defined from an initial position $x_0$ and angle $x'_0$ at location $s(0)=s_0$ and $\psi(0)= 0,$ and can be replaced in the equations of position $x$ and angle $x'$ ,
\begin{equation}
\begin{array}{ll}
x(s)=\sqrt{\epsilon}\sqrt{\beta (s)}\cos (\psi (s)+\phi)\ ,\\
x'(s) = -\frac{\sqrt{\epsilon}}{\sqrt{\beta (s)}}
{\alpha (s) \cos (\psi (s) + \phi) + \sin (\psi (s)+\phi)}\ ,
\end{array}
\label{x_x'_coord}
\end{equation}
such that they become a function of $x_0$ and $x'_0$ , as
\begin{equation}
\left ( \begin{array}{c}
x \\ x'
\end{array} \right )_{s_1}
= M
\left ( \begin{array}{c}
x\\ x'
\end{array} \right )_{s_0}\ ,
\label{eq:transport}
\end{equation}
where $M$ is the \textit{transport matrix}:
\begin{equation}
M=
\begin{pmatrix}
\sqrt{\frac{\beta}{\beta_0}}(\cos \psi+\alpha_0\sin \psi) &
\sqrt{\beta\beta_0}\sin \psi\\
\frac{(\alpha_0-\alpha)\cos\psi-
(1+\alpha\alpha_0)\sin \psi}{\sqrt{\beta\beta_0}}&
\sqrt{\frac{\beta_0}{\beta}}(\cos \psi -\alpha  \sin \psi))
 \end{pmatrix}\ .
\end{equation}
Equation (\ref{eq:transport}) is a very useful relation. The trajectory in terms of position and angle can be calculated at any point of the ring as long as the coordinates at a position $s_0$ and the so-called \textit{Twiss functions}, $\alpha$ and $\beta$, at both longitudinal positions are known.
\subsubsection{The tune}
\begin{figure}
\centering\includegraphics[width=.9\linewidth]{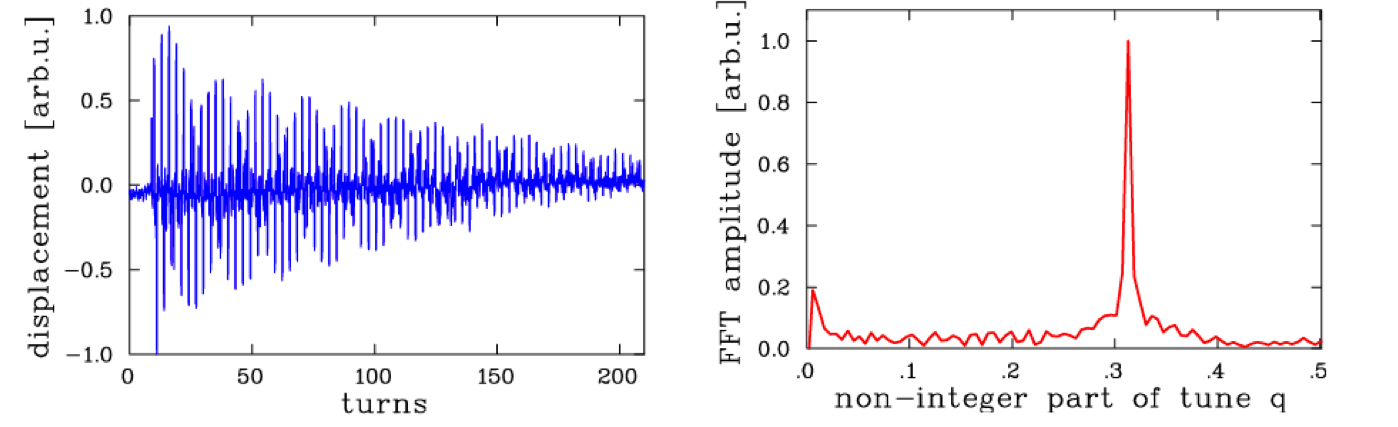}
\caption{Turn-by-turn oscillation recorded at a beam position monitor after a one-turn excitation with a kicker magnet on the left side. The fast Fourier transform spectrum of this oscillation on the right side. The amplitude peak in the spectrum indicates the tune of the oscillation.}
\label{fig:tune}
\end{figure}

Another important parameter in a circular accelerator is the so-called \textit{tune}, the number of betatron oscillations per turn
\begin{equation}
Q = \frac{\psi (L_{\mathrm{turn}})}{2\pi}=\frac{1}{2\pi}\oint \frac{\mathrm{d}s}{\beta (s)}\ .
\end{equation}
As we will see later, an exact knowledge of the tune in both transverse planes and the ability to correct the tune is of great importance for beam stability. The machine tune can be calculated from the turn-by-turn beam position data at a beam position monitor. The tune can then be obtained from the fast Fourier transform of the turn-by-turn data, as indicated in Fig. \ref{fig:tune}.

\subsubsection{Phase-space ellipse and emittance}
\begin{figure}
\centering\includegraphics[width=.5\linewidth]{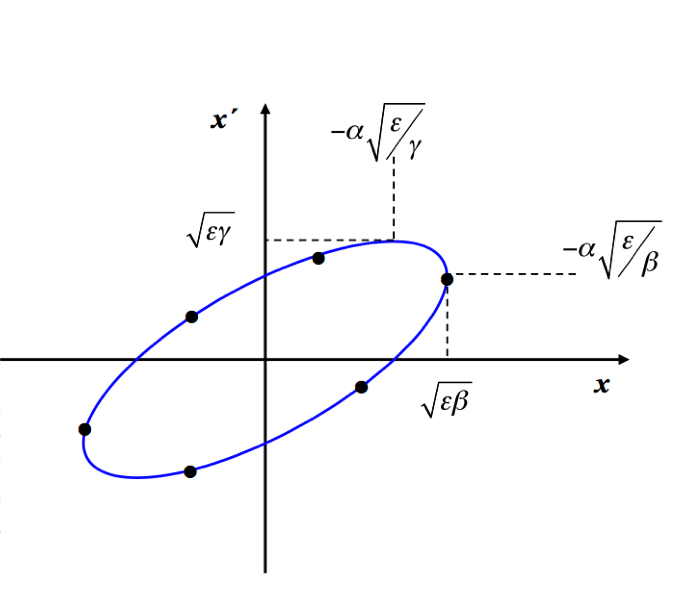}
\caption{The trajectory of a particle in phase space $x$, $x'$, turn after turn, is an ellipse. The orientation and shape is defined by the focusing properties of the lattice, whereas the area of the ellipse is an intrinsic property of the beam.}

\label{fig:ellipse}
\end{figure}

With $x$ and $x'$ in Eq. (\ref{x_x'_coord}), one can solve for $\epsilon$:
\begin{equation}
\epsilon = \gamma (s) x(s)^2+ 2\alpha(s) x(s)x'(s)+\beta(s) x'(s) ^2\ .
\label{eq:phase_space_ellipse}
\end{equation}
The result in Eq. (\ref{eq:phase_space_ellipse}) is the parametric representation of an ellipse in $x$,$x'$-space, see Fig. \ref{fig:ellipse}. The shape and orientation of the ellipse are given by the Twiss parameters, $\beta$, $\alpha$ and $\gamma$. The area of the ellipse is $A = \pi \cdot \epsilon$, which is a constant of motion according to \textit{Liouville's theorem}. Therefore, $\epsilon$ is also constant, and is called the \textit{Courant--Snyder invariant}. The area of the ellipse is an intrinsic property of the beam and cannot be changed by the focusing properties of the machine.

\begin{figure}
\centering\includegraphics[width=.5\linewidth]{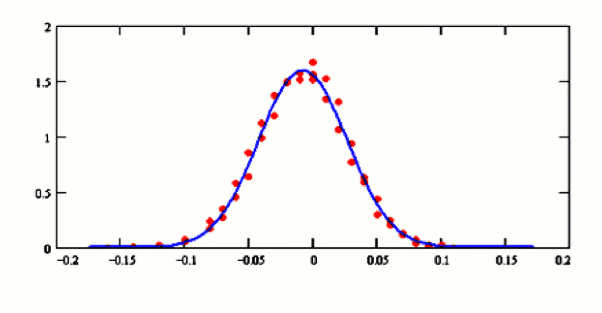}
\caption{Transverse profile measurements of a particle beam (red dots), using a wire scanner, and  Gaussian fit (blue line).}
\label{fig:wirescan}
\end{figure}

Typically, the particles in an accelerator have a Gaussian particle distribution in position and angle. The distribution in position in the horizontal plane, for example, follows the well-known relation
\begin{equation}
\rho (x) = \frac{N}{\sqrt{2\pi}\sigma_x}\cdot \mathrm{e}^{
-\frac{x^2}{2\sigma_x ^2}}\ .
\end{equation}
An example of a transverse profile measurement with a wire scanner is shown in Fig. \ref{fig:wirescan}.
The emittance $\varepsilon$ of a beam of many particles corresponds to the ellipse in phase space that contains $68.3\%$ of the particles, such that the standard deviation of a Gaussian distribution corresponds to
\begin{equation}
\sigma_x = \sqrt{\varepsilon \beta_x}\ .
\end{equation}
The beam emittance is an invariant, since the ellipse areas in phase-space are invariant. It shrinks, however, during acceleration, as will be shown next.

\textit{Liouville's} theorem from Hamiltonian mechanics states that the volume in phase space is conserved for the canonical variables $p$ and  $q$, where $q$ is typically a position coordinate and $p$ the mo\-mentum:
\begin{equation}
\int p\, \mathrm{d}q = \mathrm{const}\ .
\label{eq:liouville}
\end{equation}
For our discussion, $q$ can be written as $x$ and $p=\gamma m v = m c\gamma \beta_x$, where $\beta_x = v_x/c$.
The angle $x'$ can be transformed into
\begin{equation}
x' = \frac{\mathrm{d}x}{\mathrm{d}s}=\frac{\mathrm{d}x}{\mathrm{d}t}\frac{\mathrm{d}t}{\mathrm{d}s}=
\frac{\beta_x}{\beta}\ ,
\end{equation}
such that \textit{Liouville's} theorem from Eq. (\ref{eq:liouville}) can be rewritten as
\begin{equation}
\int p\, \mathrm{d}q = mc\int \gamma \beta_x \mathrm{d}x =
mc\gamma\beta \int x'\mathrm{d}x = \mathrm{const}\ .
\label{eq:liouville_1}
\end{equation}
During acceleration, $\gamma$ and $\beta$ increase. For the left-hand side of Eq.\ (\ref{eq:liouville_1}) to remain constant, the area in phase space and therefore also the emittance has to decrease proportionally with $1/(\beta\gamma)$. The beam size therefore shrinks during acceleration:
\begin{equation}
\varepsilon = \int x'\mathrm{d}x\propto \frac{1}{\beta\gamma}\ .
\end{equation}

\subsection{Longitudinal plane}
Circular accelerators allow for multiple application of the same RF accelerating voltage to increase the particle energy. The energy gain per turn for a sinusoidal RF voltage is
\begin{equation}
\Delta E = eV \sin \phi = eV \sin \omega_{\mathrm{RF}} t\ .
\end{equation}
The synchrotron has a fixed orbit and bending radius and a magnetic field that increases synchronously with the beam energy. A synchronous RF phase of the RF field exists, for which the energy gain of the particles fits the increase of the magnetic field.  A particle that arrives turn after turn at the same phase, $\phi = \phi_s = \mathrm{const}$, with respect to the RF field is called a \textit{synchronous particle}. For the acceleration to work, the RF frequency must be locked to the beam revolution frequency,
\begin{equation}
\omega_{\mathrm{RF}} = h\omega_{\mathrm{rev}}\ ,
\end{equation}
where $h$ is an integer and is called the harmonic number.
The energy gain per turn for the synchronous particle is $\Delta E = eV \sin \phi_s$. With $E^2 = E_0^2 + p^2 c^2 \rightarrow \Delta E = v \Delta p$ and $v = 2\pi R/T_{\mathrm{turn}}$, the energy gain can be written as
\begin{equation}
2\pi R \frac{\mathrm{d}p}{\mathrm{d}t}=q\cdot V\cdot \sin \phi_s\ .
\end{equation}
Thus, the stable phase and voltage are changed during acceleration. In the LHC, for example, the energy ramp takes more than \Unit{15}{min}, and the stable phase is close to $180^{\circ}$. The total energy gain per turn is only about 500\UkeV.

\subsubsection{The principle of phase stability}
As not all particles will go through the accelerating gap at exactly the same time, not all particles will receive the same energy gain; therefore, not all particles will have the same energy. The particles of a beam whose energy is distributed around a mean energy are all accelerated as long as the synchronous phase is chosen adequately (and the energy differences of the different particles are not too large). This is due to the \textit{principle of phase stability}.

\begin{figure}
\centering\includegraphics[width=.9\linewidth]{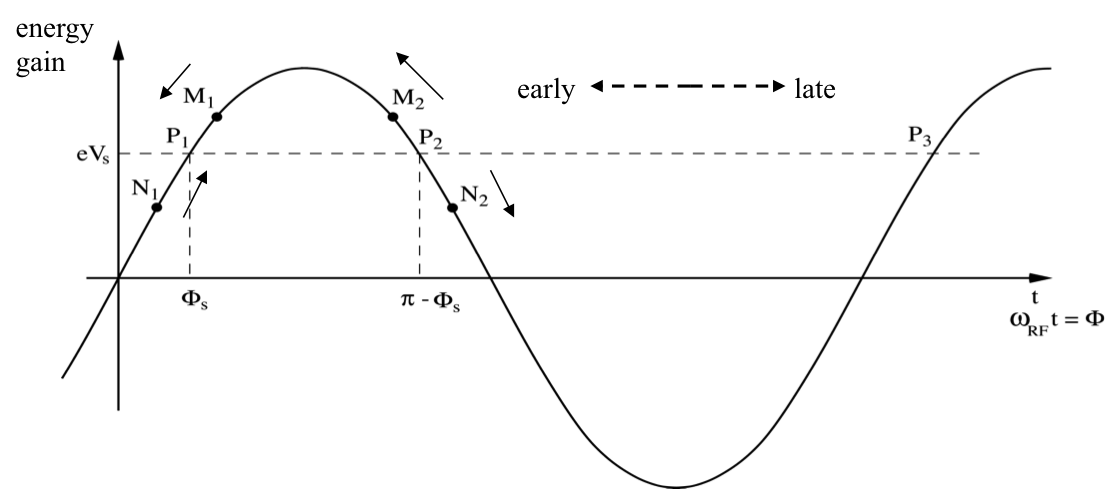}
\caption{The principle of phase stability (Courtesy of F. Tecker)}
\label{fig:phaseStability1}
\end{figure}

Let us assume a group of non-relativistic particles, where the energy increase is still transferred into velocity increase. The particles $P_1$ and $P_2$ in Fig.\ \ref{fig:phaseStability1} are two different synchronous particles; they will see the same energy gain turn after turn. The particles $N_1$, $M_1$, $N_2$ and $M_2$ represent the particle distribution around the two synchronous particles. The energy gain as a function of time for the different particles coming from the sinusoidal RF field is also shown. The particle $N_1$ had a larger energy than $P_1$ in the previous turn. It arrived earlier this turn and will get less energy than $P_1$. The particle $M_1$ on the other hand had less energy than $P_1$ and arrived later than the synchronous particle. It will get more energy this time and will move closer to the synchronous particle. It will therefore stay synchronous with acceleration.

The situation is different for the particles around synchronous particle $P_2$. Particle $M_2$ arrived earlier, as it had more energy than $P_2$ and will get even more energy during this passage through the accelerating gap. It will move away from $P_2$. Particle $N_2$ was too slow and will become even slower this time round. These particles will not stay synchronous with the acceleration and the changing magnetic field and will be lost in the vacuum chamber. The two synchronous phases for $P_1$ and $P_2$ are not equivalent. The synchronous phase has to be chosen adequately.

If a particle is shifted in momentum, it will run  on a different orbit with a different length. The par\-ameter \textit{momentum compaction}, $\alpha$, gives the relative orbit length change for a given relative momentum change:
\begin{equation}
\alpha = \frac{\mathrm{d}L / L}{\mathrm{d}p / p}\ .
\end{equation}
The particle will also have a different velocity and hence a different revolution frequency. The \textit{slippage factor} parameter, $\eta$, gives the relative revolution frequency change for a given momentum change. $\eta$ depends on the \textit{momentum compaction} as
\begin{equation}
\eta = \frac{\mathrm{d}f_{\mathrm{rev}}/f_{\mathrm{rev}}}{\mathrm{d}p/p} = \frac{1}{\gamma^2}-\alpha \ ,
\end{equation}
where $\gamma$ is the relativistic gamma. The energy corresponding to $\gamma =\gamma_\mathrm{t}= 1/\sqrt{\alpha}$ divides the longitudinal motion into two regimes. The energy $\gamma_\mathrm{t}$ is called the \textit{transition energy}.  Below transition energy ($\gamma <~\gamma_\mathrm{t}$,\ $\eta >0$), higher momentum corresponds to a higher revolution frequency. Above transition energy ($\gamma >~\gamma_\mathrm{t}$,\ $\eta < 0$), higher momentum leads to a lower revolution frequency. Below transition energy, an energy increase still leads to a velocity increase. Above the transition energy, where $v\approx c$, the velocity stays roughly constant and the increase in momentum just leads to an increase in path length. In addition, transition crossing during acceleration makes the previously stable synchronous phase unstable. In Fig. \ref{fig:phaseStability1}, the synchronous particle $P_1$ has a correct stable phase below transition energy; above the transition energy, the synchronous phase of $P_2$ has the stable phase. The moment of transition during acceleration is delicate. The RF system needs to make a rapid phase change, a \textit{phase jump}, when crossing the transition.

\subsubsection{The RF bucket and RF acceptance}
\begin{figure}
\centering\includegraphics[width=.5\linewidth]{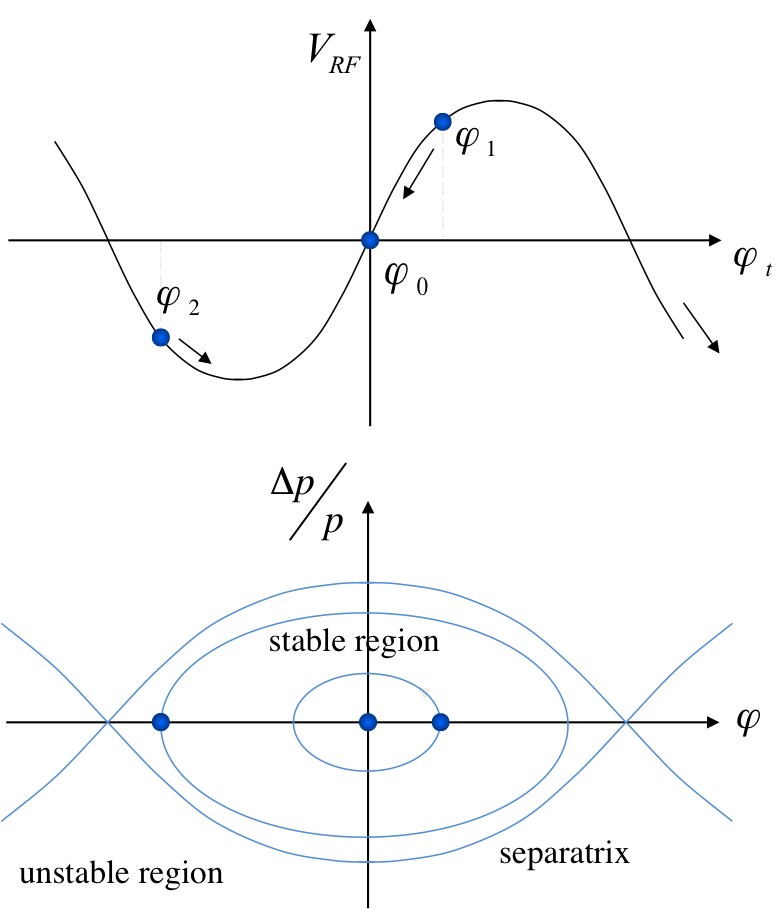}
\caption{The longitudinal motion in the upper plot follows the trajectory in phase-space in the lower plot. The separatrix defines the limit of stable motion (Courtesy of F. Tecker.)}
\label{fig:phase_space_long}
\end{figure}

\begin{figure}
\centering\includegraphics[width=.5\linewidth]{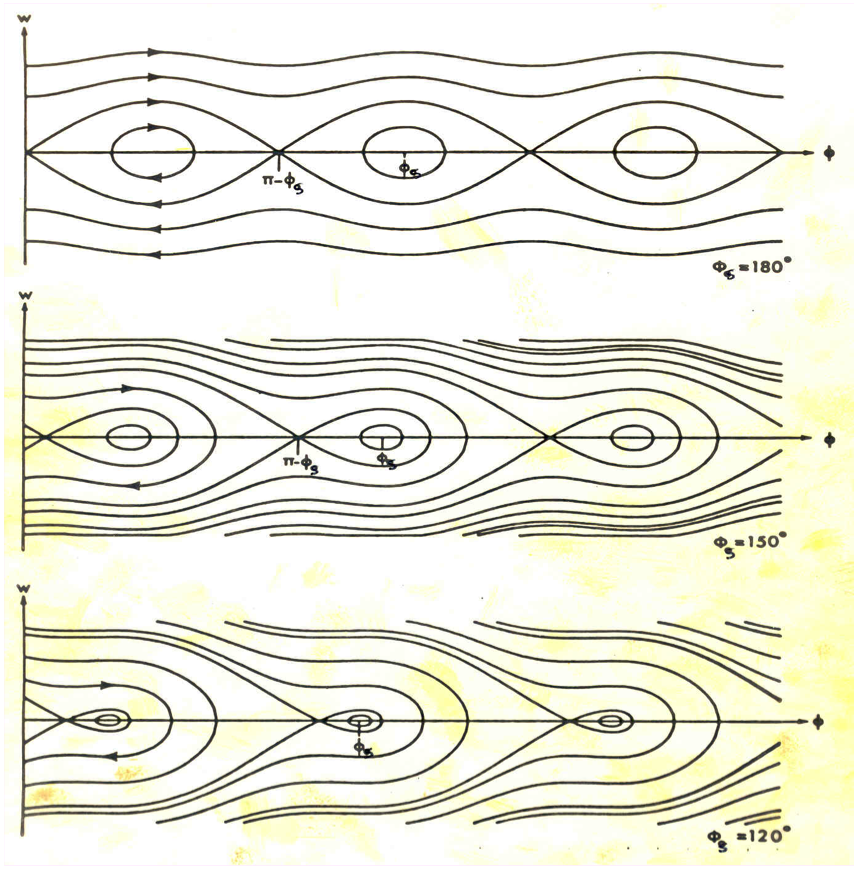}
\caption{The buckets around the ring shrink during acceleration when the synchronous phase is moved towards $90^{\circ}$ (Courtesy of F. Tecker.)}
\label{fig:RF_acceptance}
\end{figure}

As in the transverse plane, the particles are oscillating in the longitudinal plane. The particles keep oscillating around the stable synchronous particle varying phase and $\mathrm{d}p/p$, see Fig.\ \ref{fig:phase_space_long}.
The separatrix defines the region of stable motion, the so-called \textit{bucket}. The entire particle distribution needs to fit into the bucket, to avoid particle losses. The bucket area, called the \textit{RF acceptance}, is measured in electronvolts in $\Delta E$--$\Delta t$ space, which is equivalent to $\Delta p/p$--$\Delta \phi$ space. The number of buckets around the ring corresponds to the harmonic number $h$. The bucket area is largest when the synchronous phase is $0^{\circ}$,  or $180^{\circ}$, where the beam is not accelerated. For acceleration, the synchronous phase has to move towards $90^{\circ}$ and the buckets become smaller, see Fig.~\ref{fig:RF_acceptance}. The RF acceptance increases with RF voltage, however. \textit{RF acceptance} plays an important role for losses created by RF capture and stored beam lifetime. During bucket-to-bucket transfer from one machine to another, the bunches might arrive with small momentum and phase errors. If the RF acceptance is too small, part of the injected bunch ends up outside the RF acceptance. This part of the beam will not be accelerated with the rest of the beam when the momentum increases, and will be lost on the vacuum chamber.
\subsubsection{Synchrotron oscillations}
To describe the motion of the particles in the longitudinal plane, their coordinates are expressed with respect to the coordinates of the synchronous particle.  Let us call the synchronous particle $P_s$;  a particle $P$ has a phase difference with respect to the synchronous particle of
\begin{equation}
\Delta\phi = \phi - \phi_s\ ,
\end{equation}
and $P$ will also have a different revolution frequency:
\begin{equation}
\begin{aligned}
\frac{\mathrm{d}\Delta\phi}{\mathrm{d}t}& = - 2\pi h \Delta f_{\mathrm{rev}}\ ,\\
\frac{\mathrm{d}^2\Delta\phi}{\mathrm{d}t^2}& = - 2\pi h \frac{\mathrm{d} \Delta f_{\mathrm{rev}}}{\mathrm{d}t}\ .
\label{eq:towards_synch_osc_1}
\end{aligned}
\end{equation}
When the particles cross the RF cavity, the momentum increase of the two particles will be different:
\begin{equation}
\begin{aligned}
2\pi R \frac{\mathrm{d}p_s}{dt}& = q\cdot V \cdot \sin \phi_s\ ,\\
2\pi R \frac{\mathrm{d}p}{\mathrm{d}t}& = q\cdot V \cdot \sin \phi\ ,\\
2\pi R \frac{\mathrm{d}\Delta p}{\mathrm{d}t}& = q\cdot V \cdot \sin \phi - q\cdot V \cdot \sin \phi_s\ .
\end{aligned}
\label{eq:towards_synch_osc_2}
\end{equation}
Using the definition of the slippage factor,
\begin{equation*}
\eta = \frac{\mathrm{d}f_{\mathrm{rev}}/f_{\mathrm{rev}}}{\mathrm{d}p/p} = \frac{\Delta f_{\mathrm{rev}}/f_{\mathrm{rev}}}{\Delta p/p_s}\ ,
\end{equation*}
Eq. (\ref{eq:towards_synch_osc_1}) can be transformed to
\begin{equation}
\frac{\mathrm{d}^2\Delta\phi}{\mathrm{d}t^2}= - 2\pi h \frac{\mathrm{d} \Delta f_{\mathrm{rev}}}{\mathrm{d}t}=
-\frac{2\pi \eta h f_{\mathrm{rev}}}{p_s}\frac{\mathrm{d}\Delta p}{\mathrm{d}t}\ .
\end{equation}
Using Eq.\ (\ref{eq:towards_synch_osc_2}) for $\mathrm{d}\Delta p/\mathrm{d}t$, the second-order non-linear differential equation describing the synchrotron motion is obtained:
\begin{equation}
\frac{\mathrm{d}^2\Delta \phi}{\mathrm{d}t^2}+\frac{\eta\cdot f_{\mathrm{RF}}}{R\cdot p_s}q\cdot V(\sin \phi - \sin \phi_s) = 0\ .
\label{eq:synchrotron_equation}
\end{equation}
For small amplitude oscillations, with small phase deviations from the synchronous particle,  the term ($\sin \phi - \sin \phi_s$) can be written as
\begin{equation}
\sin \phi - \sin \phi_s = \sin(\phi_s + \Delta \phi) - \sin \phi_s\cong \cos\phi_s \Delta \phi\ ,
\end{equation}
and Eq. (\ref{eq:synchrotron_equation}) can be linearized to an equation of an undamped resonator with resonant frequency $\Omega_s$,
\begin{eqnarray}
\frac{\mathrm{d}^2 \Delta\phi}{\mathrm{d }t^2}+ \left[\frac{\eta f_{\mathrm{RF}} \cos \phi_s}{Rp_s} q V \right]\Delta \phi&=& \frac{\mathrm{d}^2 \Delta \phi}{\mathrm{d} t^2}+ \Omega_s ^2 \Delta \phi = 0\ ,\\
\Omega_s &=& \sqrt{\frac{\eta f_{\mathrm{RF}} \cos \phi_s}{Rp_s} q V}\ .
\end{eqnarray}
This resonant frequency, $\Omega_s$, is called the \textit{synchrotron frequency}. The periodic motion is stable if the expression under the square root of the definition of the synchrotron frequency is larger than zero. This is the case if $\eta \cdot \cos \phi_s > 0$. The necessary conditions for the synchronous phase follow from this requirement.
Below transition, the condition for the stable synchronous phase is
\begin{equation}
\gamma \leq \gamma_{\mathrm{tr}} \Rightarrow \eta \geq 0\Rightarrow \cos \phi_s \geq 0 \Rightarrow \phi_s \in [0,\ \pi/2]\ .
\end{equation}
Above transition, the condition for the stable synchronous phase becomes
\begin{equation}
\gamma \geq \gamma_{\mathrm{tr}} \Rightarrow \eta \leq 0 \Rightarrow \cos \phi_s \leq 0 \Rightarrow \phi_s \in [\pi/2,\pi]\ .
\end{equation}

\subsubsection{Dispersion}
From this discussion on the behaviour of the particles in the longitudinal plane, it is now clear that not all particles have the same momentum. In fact, a bunch contains a distribution of $\Delta p/p$. The typical momentum spread is of the order of $\mathrm{d}p/p \approx 10^{-3}$. This has been neglected so far in the discussion of transverse motion. Including this fact turns the homogeneous equations of motion in Eq. (\ref{eq:horizontal_diff_eq}) into the inhomogeneous equation.
\begin{equation}
x''+ x\left(\frac{1}{\rho^2}-k\right) = \frac{\Delta p}{p}\frac{1}{\rho}\ .
\end{equation}
The general solution to this equation is the sum of solution to the homogeneous equation $x_\mathrm{h}(s)$ and a solution that fulfils the inhomogeneous equation $x_\mathrm{i}(s)$ with  $x(s) = x_\mathrm{h}(s) +x_\mathrm{i}(s)$. The \textit{dispersion} is then defined as
\begin{equation}
D(s) = \frac{x_\mathrm{i}(s)}{\Delta p/p}\ .
\end{equation}
The \textit{dispersion} is the trajectory an ideal particle would have with $\Delta p/p = 1$. The trajectory of any particle is the sum of  $x_{\beta} (s)$ plus \textit{dispersion} $\times$ momentum offset.
$D(s)$ is just another trajectory and will therefore be subject to the focusing properties of the lattice. For a particle with momentum offset, the equation for calculating the coordinates $x$, $x'$ at any location of the ring becomes
\begin{equation}
\begin{pmatrix}
x \\ x'
\end{pmatrix}_{s_1}
= M
\begin{pmatrix}
x\\ x'
\end{pmatrix}_{s_0}
+\frac{\Delta p}{p}
\begin{pmatrix}
D\\ D'
\end{pmatrix}_{s_1}\ .
\end{equation}
\textit{Dispersion} also has an effect on the size of the beam. At a given place in the ring, the beam size depends on $\beta(s)$ and $D(s)$, together with the momentum spread of the beam, $\Delta p/p$:
\begin{equation}
\sigma = \sqrt{\beta \varepsilon +D^2 \left(\frac{\Delta p}{p}\right)^2}\ .
\end{equation}

\section{Beam loss mechanisms}
The tune, the number of betatron oscillations per turn, was introduced in Section 2.1.3. The choice of tune, and hence the focusing properties of the lattice, have important implications for the stability of motion in the presence of linear magnetic field errors and non-linear fields.
\subsection{Dipole field errors}
If the magnetic centre of a quadrupole magnet is not perfectly aligned transversely with the design orbit, or if dipole field errors are present, orbit perturbations around the ring are generated. The orbit at a location $s$ is the average trajectory over many turns at that location. With a field error $\Delta x'$ at location $s_0$ the orbit at location $s$ changes according to
\begin{equation}
x(s) = \frac{\Delta x'}{2}\cdot \sqrt{\beta (s_0)\beta (s)}
\frac{\cos (\pi Q -\psi_{s_0\rightarrow s})}{\sin (\pi Q)}\ .
\label{eq:orbit_distortion}
\end{equation}
The effect of the error, $\Delta x'$, will be large at locations with large $\beta(s)$ and depends on the phase advance between the error location and the orbit location of interest. Also, the larger $\beta(s_0)$ is at the error location, the larger will be the effect of the error around the ring. The other important lesson to be drawn from Eq.~(\ref{eq:orbit_distortion}) is the dependence on the tune. With the term $\sin (\pi Q)$ in the denominator, the orbit response around the ring for any dipole error diverges for $Q = N$, where $N$ is an integer.
\subsection{Gradient errors}
Gradient errors will lead to a change in the tune and the beta functions around the ring.
The tune change can be calculated by evaluating the distorted one-turn matrix with a small field error $\Delta k$ over a distance $l$, which might be the length of a magnet. The one-turn matrix is the transport matrix from location $s \to s+L = s$, where $L$ is the circumference of the ring:
\begin{equation}
\begin{aligned}
M_{{\mathrm{turn}}_{\mathrm{dist}}}& =
\begin{pmatrix}
\cos 2\pi Q+\alpha\sin 2\pi Q &
\beta\sin 2 \pi Q\\
-\gamma \sin 2\pi Q &  \cos 2\pi Q -\alpha  \sin 2\pi Q
\end{pmatrix} \\
& =
\begin{pmatrix}
1 & 0\\
- \Delta k l&  1
 \end{pmatrix}
\cdot
\begin{pmatrix}
\cos 2\pi Q_0+\alpha\sin 2\pi Q_0 &
\beta\sin 2 \pi Q_0\\
-\gamma \sin 2\pi Q_0 &  \cos 2\pi Q_0 -\alpha  \sin 2\pi Q_0

\end{pmatrix}\ ,
\end{aligned}
\end{equation}
where $Q = Q_0+\Delta Q$. With $\mathrm{Tr}(M_{\mathrm{turn_{dist}}}) = \mathrm{Tr}(M_{\mathrm{error}}\cdot M_{\mathrm{turn}})$, the tune change evaluates to
\begin{equation}
\Delta Q = \frac{1}{4\pi}\beta \Delta k \cdot l\ .
\label{eq:tune_change}
\end{equation}
The larger the beta function at the location of the gradient error, the larger the tune change. The relative beta function change, the so-called \textit{beta-beat}, due to the gradient error $\Delta k$ is
\begin{equation}
\frac{\Delta \beta (s)}{\beta (s)} = -\frac{1}{2 \sin (2\pi Q)}\beta (s_{0})\cos \left[2\left(\psi (s_0)-\psi (s)\right)-2\pi Q\right]\cdot \Delta k\cdot l\ .
\end{equation}
As discussed earlier, the beta function is related to the size of the beam with $\sigma = \sqrt{\beta \varepsilon}$. The beta functions, and hence the beam sizes, diverge with any gradient error if the tune $Q = N,\ N/2$, where $N$ is an integer, owing to the term $\sin 2\pi Q$ in the denominator.
\subsection{Non-linear imperfections}
In the Taylor series expansion of the magnetic field for the derivation of the equation motion in the transverse plane, the higher-order components in $x$ or $y$ were neglected. Higher-order fields might, however, be present, owing to non-perfect dipole and quadrupole magnets, or they might be introduced on purpose, to stabilize the beam. The magnetic field of multiple of order $n$ is
\begin{equation}
B_y(x,y) +\mathrm{i}\cdot B_x(x,y) = (B_n(s)+\mathrm{i} A_n(s))\cdot (x+\mathrm{i}y)^n\ ,
\end{equation}
where $B_n(s)$ are the normal coefficients and $A_n(s)$  are the skew coefficients from the Taylor series expansion,
\begin{equation}
\begin{aligned}
B_n (s)& =\frac{1}{(n)!}\frac{\partial^{n} B_y}{\partial x^{n}}\ , \\
A_n (s)& =\frac{1}{(n)!}\frac{\partial^{n} B_x}{\partial x^{n}}\ .
\end{aligned}
\end{equation}
For example, $\partial^{2} B_y/\partial x^{2}$  and $\partial^{3} B_y/\partial x^{3}$ are                                  the sextupole and octupole component, respectively. In the presence of non-linear fields, the equation of motion becomes a non-linear differential equation. An example for the horizontal plane would be
\begin{equation}
\frac{\mathrm{d}^2 x}{\mathrm{d}s^2}+K(s) \cdot x = \frac{F_x}{v\cdot p}\ ,
\end{equation}
where $F_x$ is the Lorentz force from the non-linear magnetic field.

\begin{figure}
\centering\includegraphics[width=.5\linewidth]{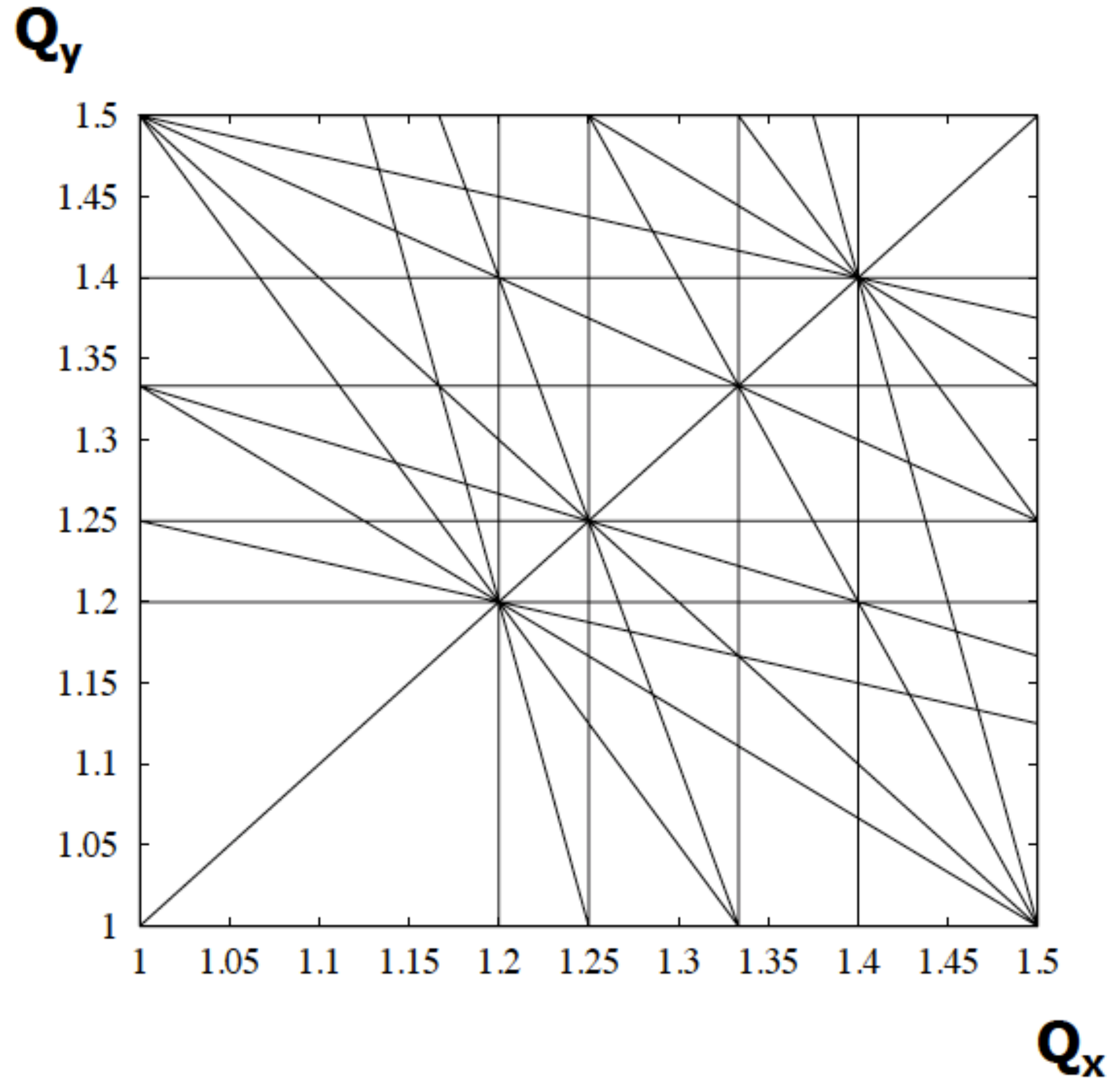}
\caption{The tune diagram indicates forbidden resonance lines. The chosen machine tunes in the horizontal and vertical plane have to be far from these lines.}
\label{fig:tuneDiagram}
\end{figure}

It was mentioned earlier that the motion becomes unstable for $Q = N,\ N/2$ in the case of quadrupole field errors. It can be shown for sextupole perturbations that amplitudes increase for $Q~=~N,\ N/3$; and for octupole perturbations that amplitudes increase for $Q = N,\ N/2,\ N/4$. In general, the machine tune has to be chosen such that it does not fulfil the condition
\begin{equation}
nQ_x + mQ_y = N\ ,
\end{equation}
where $n$, $m$ and $N$ are small integers. The forbidden tunes are often summarized in the so-called tune diagram as \textit{resonance lines}. The resonance lines with the lowest order are the most dangerous ones. An example of a tune diagram with low-order resonance lines is shown in Fig. \ref{fig:tuneDiagram}.
\subsubsection{Sextupole fields}
\begin{figure}
\centering\includegraphics[width=.9\linewidth]{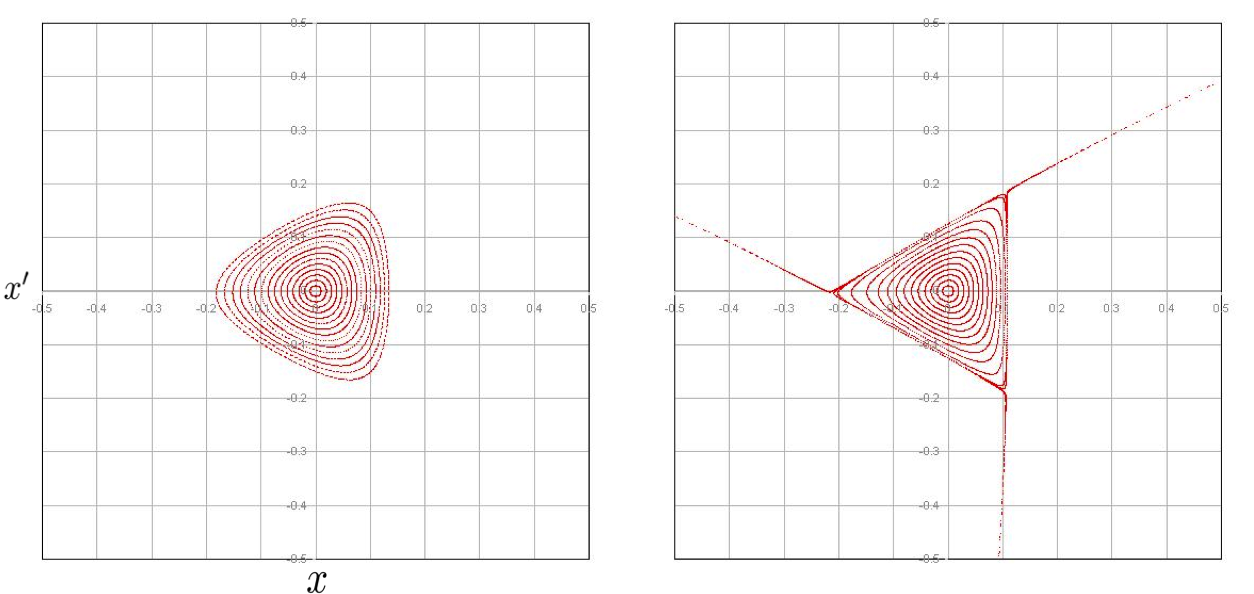}
\caption{The trajectories in phase space $x, x'$ with sextuple fields in the ring. The size of the stable area within the triangle is a function of the distance of the tune from the third-order resonance.}
\label{fig:sextupole_phaseSpace}
\end{figure}

The example of the sextuple fields and their effect on particle motion will be briefly discussed to further explain the effect of non-linear fields. The resulting fields of a sextuple coil configuration are
\begin{align*}
B_x &= \tilde{g}xy\ ,\\
 B_y&= \frac{1}{2}\tilde{g}\left(x^2 - y^2\right)\ .
\end{align*}
The sextupole fields generate a gradient in both planes, rising linearly with the offset in $x$  according to
\begin{align*}
\frac{\partial B_x}{\partial y}=\frac{\partial B_y}{\partial x}=\tilde{g}x\ .
\end{align*}
The equations of motion in the presence of the sextupole field become
\begin{align}
x''+K_x(s) & = - \frac{1}{2} m_{\mathrm{sext}}(s) (x^2 - y^2)\ ,\\
y''+ K_y (s) & =  m_{\mathrm{sext}}(s) xy\ ,
\end{align}
where $m_{\mathrm{sext}}$ is the normalized sextuple strength with $m_{\mathrm{sext}} = \tilde{g}/(p/e)$.
The effect of a single sextuple around the ring on the particle trajectories can be simulated by adding a sextuple \textit{kick} at the location of the sextuple,
\begin{align*}
\Delta x' & = -\frac{1}{2}m_{\mathrm{sext}} l (x^2-y^2)\ ,\\
\Delta y'& = m_{\mathrm{sext}} l xy\ ,
\end{align*}
where $l$ is the length of the sextuple. As a result of these kicks, the phase-space trajectories become more and more distorted for larger amplitudes and are no longer elliptical, see Fig.\ \ref{fig:sextupole_phaseSpace}. The motion becomes unstable, meaning that the amplitudes of the particles become larger and larger after each turn. The size of the stable area within the triangle in phase space is proportional to  $(Q - \frac{p}{3})/(m_{\mathrm{sext}})$, where $p$ is an integer.

\subsubsection{Chromaticity}
The normalized quadrupole gradient was defined as $k = g/(p/e)$. The different particles in a beam have a distribution of momenta around the ideal momentum $p_0$. For a given particle, $p=p_0+ \Delta p$, and the normalized gradient for this particle is
\begin{equation*}
k=\frac{eg}{p_0+\Delta p}\approx \frac{e}{p_0}\left(1-\frac{\Delta p}{p_0}\right)g=k_0+\Delta k\ ,
\end{equation*}
where the gradient error is
\begin{equation*}
\Delta k = - \frac{\Delta p}{p_0}k_0\ .
\end{equation*}
As discussed already, gradient errors result in tune changes. Thus, particles with a different momentum $p$ distributed around $p_0$ will all have different tunes.
Using Eq. (\ref{eq:tune_change}), the tune change for a particle of a given momentum offset $\Delta p/p_0$ is
\begin{equation}
\Delta Q = \frac{1}{4\pi}\beta \Delta k \cdot l = - \frac{1}{4\pi}\frac{\Delta p}{p_0}k_0\beta l\ .
\label{eq:quadrupole_chroma}
\end{equation}
The parameter \textit{chromaticity} is defined as the ratio of the tune change for a given relative momentum change:
\begin{equation}
\Delta Q = Q' \frac{\Delta p}{p}\ ,
\end{equation}
and, from Eq. (\ref{eq:quadrupole_chroma}),  the \textit{chromaticity} of a synchrotron equates to
\begin{equation}
Q' = -\frac{1}{4\pi}\oint k(s) \beta (s) \mathrm{d}s\ .
\end{equation}
\textit{Chromaticity} is created in the vertical and horizontal planes by the quadrupole fields. Together with the beam momentum spread, it indicates the size of the tune spot in the tune diagram, which will no longer be a single point. The natural chromaticity in the LHC, for example, is \Unit{250}{units}. With a typical momentum spread of $\Delta p/p = 0.2 \times 10^{-3}$ and fractional injection tune $Q_x = 0.28$, the particles would have tunes between $Q_x=0.26$ and $Q_x=0.33$. The particle distribution would cross several dangerous resonance lines, leading to beam loss. Chromaticity, therefore, has to be corrected. The sorting of the particle amplitudes in the horizontal plane due to dispersion $x_D(s)= D(s) \frac{\Delta p}{p}$ is used for this purpose. Sextupole magnets are placed at locations with large dispersion $D_x$. The resulting gradient at the sextupole location depends on the particle amplitude in the horizontal plane, and the sextupole strength is chosen such that the chromaticity in both planes is adjusted to a suitable value. $Q'= 0$ is, however, not necessarily desirable, owing to so-called \textit{collective effects}.
\subsection{Collective effects}
\begin{figure}
\centering\includegraphics[width=.5\linewidth]{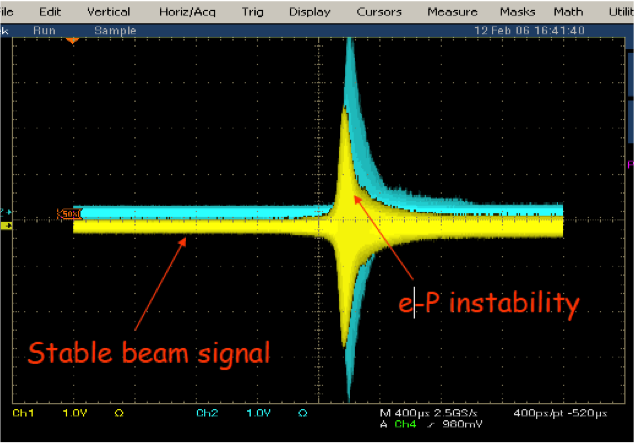}
\caption{Beam position measurement at a beam position monitor turn-by-turn. During the instability, the trajectory amplitude grows exponentially until the beam loses intensity and stabilizes itself.}
\label{fig:instability}
\end{figure}

Collective effects can cause beam instabilities, emittance blow-up and beam loss. A typical example of the turn-by-turn trajectory during a beam instability is shown in Fig. \ref{fig:instability}. There are three main categories of \textit{collective effect}:
\begin{itemize}
\item[]  \textbf{Beam--self: }The beam interacts with itself through \textit{space-charge},  causing a tune spread pro\-portional to  $1/(\beta^2\gamma^3)$. This effect is one of the main brightness limitations in low-energy machines.
\item[] \textbf{Beam--beam: }Beams interact with each other at or close to the collision point in colliders or they interact with ambient electron clouds. Colliding beams produce large tune spreads and tune shifts, owing to head-on and long-range collisions.
\item[] \textbf{Beam--environment: }These are impedance-related instabilities. The beam induces electro\-magnetic fields in the accelerator environment, such as in the vacuum chambers. These so-called \textit{wake fields} can act back on the trailing beam. The Fourier transform of the \textit{wake field} is called the impedance. Energy is lost, owing to the \textit{wake field}. If the energy remains  trapped, it can lead to component heating. If the energy is transferred to the trailing beam, it can cause beam instabilities.
\end{itemize}

 It is possible to stabilize the beam under the influence of collective effects, using such mitigative means as transverse feedback with sufficient bandwidth or the introduction of tune spread through non-zero chromaticity, octupole fields, or the head-on beam--beam effect in colliders for \textit{Landau damping}. A coherent oscillation at a frequency within the beam frequency spread is generally damped.
\subsubsection{Space-charge effect}
The space-charge effect is the simplest and most fundamental of all collective effects. It will be treated in a simple approximation as an example for collective effects associated with \textit{direct space-charge}.

\begin{figure}
\centering\includegraphics[width=.3\linewidth]{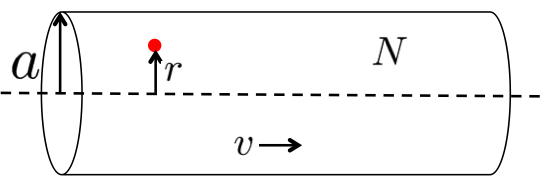}
\caption{For a cylindrical beam, the space-charge force on a particle at distance $r$ from the beam centre can easily be easily calculated.}
\label{fig:space_charge}
\end{figure}

Let us assume that  the beam is a long uniformly charged cylinder of current $I$, as indicated in Fig.~\ref{fig:space_charge}. The force exerted on a particle by the surrounding beam  at a distance $r$ from the beam centre is:
\begin{equation}
F_{r} = F_E + F_B = \frac{eI}{2\pi c\beta \varepsilon_0\gamma^2 a^2}r\ .
\label{eq:space_charge_force}
\end{equation}
For simplicity, it can be assumed that the particle has only a horizontal offset from the beam centre; in the Frenet--Serret coordinate system, Eq. (\ref{eq:space_charge_force}) becomes
\begin{equation}
F_{x} = \frac{eI}{2\pi c\beta \varepsilon_0\gamma^2 a^2}x\ .
\end{equation}
The influence of the space-charge force on transverse motion is derived by treating it as perturbation of the linear equation of motion, as  discussed with the non-linear fields:
\begin{align*}
x''(s)+K(s)x & =  \frac{F_{\mathrm{SC}}}{m\gamma\beta^2 c^2}\ ,\\
x''(s) +\left ( K(s)-\frac{2r_0I}{ea^2\beta^3\gamma^3 c } \right ) x & = 0\ ,
\end{align*}
where $r_0 = e^2/(4\pi \varepsilon_0 m_0 c^2)$ is the classical particle radius. As the space-charge force is linear in $x$, it introduces a gradient error, and gradient errors lead to tune shift. The tune shift from  the space-charge around the whole ring with radius $R$ is
\begin{equation}
\Delta Q_x = \frac{1}{4\pi}\int_0^{2\pi R} \beta_x(s) \Delta K _{\mathrm{SC}} (s)\mathrm{ d}s =  - \frac{r_0RI}{e\beta^3\gamma^3 c}\left \langle \frac{\beta_x(s)}{a^2 (s)}\right \rangle\ .
\end{equation}
As $a$ is related to the transverse size of the cylindrical beam, $a^2/\beta_x$ is related to the invariant of motion, $\varepsilon_x = a^2/\beta_x$.
The space-charge tune shift can therefore be written as
\begin{equation}
\Delta Q_x = - \frac{nr_0 }{2\pi\varepsilon_x\beta^2\gamma^3}\ ,
\end{equation}
with  $I = (n e \beta c)/(2\pi R)$. The tune shift is larger if the beam has higher brightness $n/(\varepsilon_{x,y})$ and lower energy.

\begin{figure}
\centering\includegraphics[width=.5\linewidth]{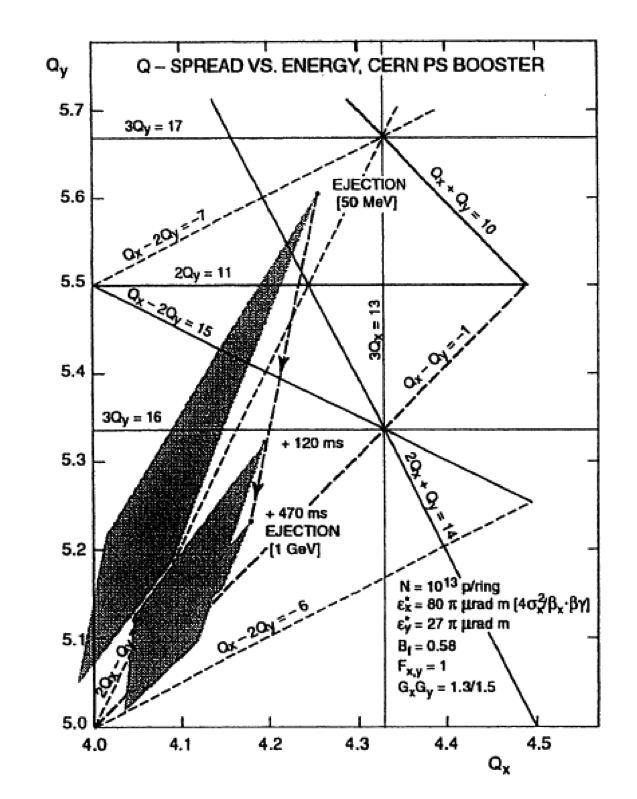}
\caption{Tune spread due to the space-charge effect for high-intensity beams in the CERN PS Booster. Injection energy, 50\UMeV; maximum extraction energy towards the CERN PS, 1.4\UGeV.}
\label{fig:spaceCharge_booster}
\end{figure}

In realistic beams, the particle density will not be uniform and the different particles will see different space-charge forces, depending on where they are in the beam. This will introduce a tune spread instead of a tune shift. An example of the tune spread due to space-charge in the CERN PS Booster is shown in Fig. \ref{fig:spaceCharge_booster}.

\section{Accidental beam loss}
Unforeseen aperture limitations can cause beam losses. The aperture limitation may be the result of misalignment of equipment or movable equipment, such as screens, girders or collimators, that is only partially retracted. Aperture limitations may also be introduced by orbit changes from quadrupole misalignment or orbit bumps. Orbit bumps might be needed for extraction systems or to introduce crossing angles in colliders. Orbit bumps might also result from corrections made as a result of false beam position monitor readings.

Fast transverse kickers, such as aperture kicker magnets, tune kicker magnets, crab cavities or transverse feedback systems, can excite significant fractions of the beam to high amplitudes within a single or a few turns. Power supply limitations or special run configurations are put in place to avoid large deflections from the aperture kicker at high intensity in the LHC. Injection kickers or extraction kickers firing asynchronously can be very dangerous for high-intensity machines. This topic is discussed further in Ref. \cite{bib:vk_bt}.

Beam losses can also originate from uncaptured beams. Uncaptured beam is generated at injection if part of the bunch is injected outside the RF acceptance. Many mechanisms can drive the beam out of the bucket after injection, for example, intra-beam scattering, beam--beam interactions or malfunctioning equipment. Malfunctioning equipment comprises noise in the phase loop, badly adjusted longitudinal blow-up or the switch-off of an RF cavity, resulting in a reduction in the total available voltage and hence reduction of the RF acceptance. Once uncaptured, the time taken for the particles to get lost on the vacuum chamber depends on the energy loss per turn due to synchrotron radiation, electron cloud or impedance and the so-called \textit{momentum aperture} in $\Delta p/p$ for the given mechanical aperture, dispersion and beta function.
\subsection{Powering failures}
The power supplies of the circuits powering the machine elements can fail and the magnetic field seen by the beam of the concerned elements will then decay. Failing dipole magnets will generate an orbit distortion that changes with time. If the failure is sufficiently slow (e.g.\ in the LHC, over more than 10 turns), the closed orbit formula in Eq.\ (\ref{eq:orbit_distortion}) for a time-varying field error can be used to calculate what happens to the beam,
\begin{equation}
\Delta x_{\mathrm{CO}} (s) = \frac{\sqrt{\beta(s)}}{2\sin (\pi Q)} \left ( \frac{I(t)}{I_0}-1\right)\sum^{N}_{i}\theta_i\sqrt{\beta (s_i)}\cos (\psi (s)-\psi (s_i) +\pi Q)\ ,
\end{equation}
where $N$ is the number of dipole magnets connected in series to the failing power supply. A power supply has many possible failure cases. The function $I(t)$ needs to be established. In most powering failure cases, the voltage is set to zero and the current decays as
\begin{equation}
I(t) = I_0 \mathrm{e}^{-\frac{t}{\tau}}\ ,
\end{equation}
where the time constant $\tau$ is defined by the resistance and inductance of the circuit as $\tau = L/R$. In the case of a main dipole magnet quench in the LHC, the current in the quenching magnet would decay as
\begin{equation}
I(t) = I_0 \mathrm{e}^{-\frac{t^2}{2\sigma^2}}\ ,
\end{equation}
where $\sigma$ has been found to be $\sigma = 200\Ums$.
A failing quadrupole circuit will move the tune in the tune diagram with the risk of crossing resonance lines. The beam size will also change as a result of the change of the beta functions, as discussed for gradient errors. If the orbit is not zero in the quadrupole, the failing quadrupole will also change the orbit. This discussion can be extended to cover higher-order circuits. The higher the order, the less critical the circuits tend to be.

\end{document}